\documentclass[]{aa}
\usepackage{graphics}
\begin{document}

\thesaurus{02.07.2; 02.13.1; 08.16.7 Crab; 08.16.7 Vela; 08.16.7 PSR 0540-69;
08.16.7 PSR 1509-58}
\title{Pulsars ellipticity revised}

\author{C.\,Palomba\inst{1,}\inst{2}}

\institute{Dipartimento di Fisica "G. Marconi", Universit\`a di Roma
"La Sapienza" \and Sezione INFN Roma1, p.le A. Moro 5, I-00185 Roma, Italy}
\mail{cristiano.palomba@roma1.infn.it}
\date{Received ..... / Accepted .....}

\maketitle

\begin{abstract}
We derive new upper limits on the ellipticity of pulsars whose braking
index has been measured, more
tightening than those usually given, assuming that both a gravitational torque
and an electromagnetic one act on them. We show that the measured braking 
indexes of pulsars are recovered in this model. We consider the
electromagnetic torque both constant and varying with time. At the same time
constraints on the pulsars initial period and on the amplitude of
the gravitational waves emitted are obtained. 
\end{abstract}

\section{Introduction}
From the timing measurements we deduce that radio pulsars are subject to a
systematic secular spin-down.
The standard formula for pulsar spin-down is
\begin{equation}
\dot{\Omega}= K\Omega ^n
\label{eq1}
\end{equation}
The quantity $I\dot{\Omega}$, being $I$ the star momentum of inertia with 
respect to its rotation axis, is the torque acting on the star and we 
will refer to $K$, following Allen $\&$ Horvath (\cite{horv2}), as the
``torque function''.  
$K$ and the braking index $ n$ depend on the mechanism which is at work. For instance,
$ n=3$ for pure magnetic dipole radiation and $ n=5$ for gravitational radiation.
The braking index can be calculated, at least in principle, from
the relation
\begin{equation}
n={{\Omega ~ \ddot{\Omega}}\over{{\dot{\Omega}}^2}}
\label{none}
\end{equation}
which can be very easily derived from Eq.(\ref{eq1}).
In practice, only for four pulsars the measure of the braking index 
is not dominated by the timing noise (Lyne et al. \cite{lyne1}),
(Boyd et al. \cite{boyd}),(Kaspi et al. \cite{kaspi}),
(Lyne et al. \cite{lyne2}): 
$$ Crab:~~~~~n=2.51\pm 0.01$$
$$ PSR~0540-69:~~~n=2.28\pm 0.02$$
$$ PSR~1509-58:~~~n=2.837\pm 0.001 $$
$$ Vela:~~~~~n=1.4\pm 0.2$$
All these values are less than the canonical value, $3$, which we expect 
for the emission of dipolar electromagnetic radiation. As we will see in the
next section,
different mechanisms have been proposed to explain this discrepancy. 
On the other hand, when pulsars are considered as possible sources of
gravitational radiation, the upper limit on the amplitude of the waves emitted
is calculated assuming that all the spin-down is due to the emission of 
gravitational waves, e.g. (Haensel \cite{haens}), (Giazzotto et al. \cite{giazo}). This implies, 
see Eq.(\ref{ompgw}), a braking index equal to 5. Clearly, this is not true,
at least for the pulsars whose braking index has been measured. A much more
realistic hypothesis is that in a pulsar both an electromagnetic and a
gravitational torques work to produce the observed spin-down. We will show
how from this assumption the observed braking indexes can be derived and, at 
the same time, how new more tightening upper limits on the pulsars ellipticity
are obtained. 

The plan of the paper is the following.
In Sec.\ref{em} we give the main relations relative to the electromagnetic
braking and shortly describe different proposed mechanisms which could produce
the observed braking indexes. In Sec.\ref{gw}, after the introduction of the 
relations describing the
gravitational torque, we show how the observed braking indexes can
be obtained combining the electromagnetic and gravitational torques.
In Sec.\ref{evol} we calculate, in the framework of our model,
the time evolution of various observable
quantities, in particular the braking index $n(t)$ and the angular velocity of
the pulsar $\Omega(t)$, and derive new upper limits on the ellipticity of
pulsars and then, Sec.(\ref{graw}), on the amplitude of the gravitational signals emitted.
Finally, in Sec.\ref{concl} conclusions are discussed.

\section{The electromagnetic torque}
\label{em}
As it is well known, the electromagnetic torque implies an evolution of the
pulsar rotation frequency given by the relation
\begin{equation}
\dot{\Omega}_{em}=K_{em}\Omega^{n_{em}}
\label{ompem}
\end{equation}
The electromagnetic index is $n_{em}=3$ for pure
dipolar magnetic radiation (Ostriker $\&$ Gunn \cite{ostri}), 
(Manchester $\&$ Taylor \cite{manch}), in which case
\begin{equation} 
K_{em}=-{2\over{3c^3}}{B^2\over{I}}R^6\sin^2{(\alpha)}
\label{kem}
\end{equation}
In this equation $ B$ is the strength of the magnetic field, $ R$ is the 
neutron star radius, $ \alpha $ is the angle between the rotation and the
magnetic axes, and $ I$ is the momentum of inertia with respect to the rotation
axis. An electromagnetic braking index different from 3 can be the consequence of
pulsar winds, in which particles having angular momentum are accelerated away from
the pulsar (Goldreich $\&$ Julian \cite{gold}), (Kaspi et al. \cite{kaspi})
or can be produced if non
dipolar components of the magnetic field are present, as a consequence of strong 
magnetospheric currents (Blandford $\&$ Romani \cite{roma}). 
Another possibility is that the magnetic moment, and then 
the torque applied to the star, varies in time (Blandford $\&$ Romani \cite{roma}),
(Cheng \cite{chen}). 
Muslimov $\&$ Page
(\cite{musli}) consider the time evolution of the surface magnetic field of the star, 
which, according to their model,
is very low ($ 10^8\div 10^9~G$) for a newborn neutron star and then increases 
due to the ohmic diffusion of the initially trapped inner magnetic field. 
The resulting electromagnetic braking index is given by
\begin{equation}
n_{em}=3+2\left({\dot{B}_{surf}\over{B_{surf}}}\right)\left({\Omega \over{\dot{\Omega}}}
\right)
\end{equation}
where $ B_{surf}$ is the value of the surface magnetic field at the 
magnetic pole. Clearly, if $ \dot{B}_{surf}>0$, $ n_{em}<3$.
Other models based on the secular variation of the magnetic field have been
developed, for instance, by Blandford {\em et al.} (\cite{bland}) and by
Camilo (\cite{cami}).
A different kind of models has
been proposed by Allen $\&$ Horvath (\cite{horv1}) and by Link $\&$ Epstein
(\cite{link}), based on the growth of the angle $\alpha $
between the magnetic moment and the rotation axis of the star. In this case the
electromagnetic braking index is
\begin{equation}
n_{em}=3+2{\dot{\alpha}\over{\tan{\alpha}}}\left({\Omega \over{\dot{\Omega}}}
\right)
\end{equation}
In all these cases the general expression for the observed electromagnetic braking index
is simply
\begin{equation}
n_{em}=3+{\dot{K}_{em}\over{K_{em}}}\times {{\Omega}\over{\dot{\Omega}_{em}}}
\label{dk}
\end{equation}
which is easily obtained deriving Eq.(\ref{ompem}) with respect to time.
From Eq.(\ref{kem}) we find
\begin{equation}
{\dot{K}_{em}\over{K_{em}}}={1\over{B\sin^2{\alpha}}}\left[
2\dot{B}\sin^2{\alpha}+B\sin{2\alpha}~\dot{\alpha}-B\sin^2{\alpha}{\dot{I}\over{I}}
\right]
\label{dkm}
\end{equation}
Assuming $n_{em}=const$, from Eq.(\ref{dk}) 
we find 
\begin{equation}
K_{em}=K_{em,0}\times \left({\Omega\over {\Omega_0}}\right)^{n_{em}-3}
\label{kom}
\end{equation}
where $K_{em,0}$ and $\Omega_0$ are the electromagnetic torque function and the
angular velocity at time $t=0$, i.e. now.
Then, the rate of variation of the angular velocity can be written as
\begin{equation}
\dot{\Omega}_{em}=K_{em,eff}\Omega^{n_{em}}
\label{nomem}
\end{equation}
where $K_{em,eff}={K_{em,0}\over{\Omega^{n_{em}-3}_0}}$ is a
constant, which is formally equivalent to Eq.(\ref{ompem}). 
Integrating Eq.(\ref{ompem}) we obtain the pulsar ``characteristic'' time,
due to the electromagnetic braking which is:
\begin{equation}
\tau_{em}={\Omega_0\over{(1-n_{em})\dot{\Omega}_0}}\left[1-\left({\Omega_0
\over{\Omega_i}}\right)^{n_{em}-1}\right]
\label{tem}
\end{equation}
being $\Omega_i$ the initial angular velocity of the pulsar. Eq.(\ref{tem})
would give the 
true age if only the electromagnetic emission was responsible
for the pulsar spin-down. 

From Eqs.(\ref{ompem},\ref{kem}), 
the strength of the magnetic field, for $n_{em}=3$, is
\begin{equation}
B\sin{\alpha}\simeq 3.2\cdot 10^{19}\sqrt{P\dot{P}_{em}}~~G
\end{equation}
where $ \alpha $ is the angle between the magnetic and the rotation axis, 
$\dot{P}_{em}$ is the rate of variation of the pulsar period due to the electromagnetic
torque, and we have assumed $I=10^{38}~kg~ m^2$. 

\section{Electromagnetic plus gravitational torque}
\label{gw}
In this section we show that the
observed braking indexes can result from the combined action, in a 
pulsar, of the magnetic and gravitational torques.
We assume that the emission of electromagnetic and gravitational radiation 
are the only mechanisms 
acting in the pulsar. The observed spin-down rate can be expressed as
\begin{equation}
\dot{\Omega}=\dot{\Omega}_{em}+\dot{\Omega}_{gw}
\label{omp}
\end{equation}
where $\dot{\Omega}_{em}$ is given by Eq.(\ref{ompem}), or Eq.(\ref{nomem}), while 
\begin{equation}
\dot{\Omega}_{gw}=-{{32}\over 5}{G\over c^5}I\epsilon^2\Omega^5\equiv
K_{gw}\Omega^5
\label{ompgw}
\end{equation}
where $\epsilon $ is the ellipticity of the pulsar.
We note that, in analogy with Eq.(\ref{tem}), the pulsar ``characteristic''
age, due to the gravitational braking, is
\begin{equation}
\tau_{gw}=-{\Omega_0\over{4\dot{\Omega}_0}}\left[1-\left({\Omega_0
\over{\Omega_i}}\right)^4\right]
\label{tgw}
\end{equation}
In the following, we will consider the torque functions $K_{em}$ and $K_{gw}$
constant in time.

\begin{figure}
\resizebox{\hsize}{!}{\includegraphics{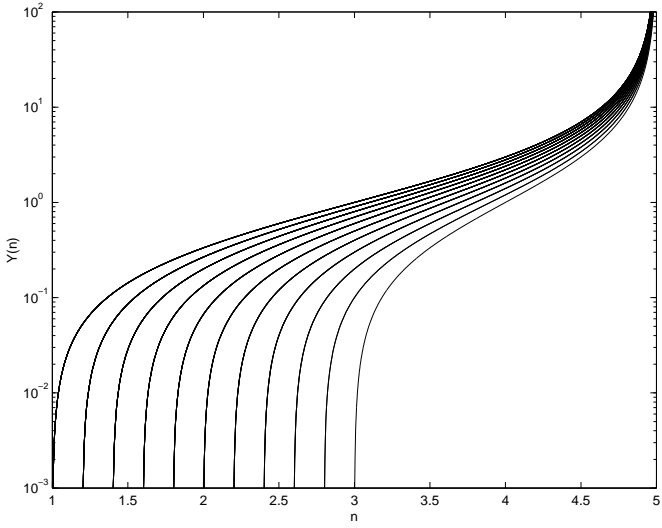}}
\caption{$Y={\dot{\Omega}_{gw}\over{\dot{\Omega}_{em}}}$ as a function
of the braking index $ n$ for different values of the electromagnetic 
braking index $ n_{em}$. $n_{em}$ varies in the range $[1,3]$ with step 0.2.}
\label{fig1}
\end{figure}

The braking index $ n$ can be obtained differentiating Eq.(\ref{omp}), 
and using Eqs.(\ref{ompem},\ref{ompgw}):
\begin{equation}
n={{\Omega\times \ddot{\Omega}}\over{\dot{\Omega}^2}}=
{{n_{em}+5Y}\over {1+Y}}
\label{n1}
\end{equation}
where we have introduced 
\begin{equation}
Y(\Omega)={\dot{\Omega}_{gw}\over{\dot{\Omega}_{em}}}={K_{gw}\over
K_{em}}\Omega^{5-n_{em}}
\label{ipsi}
\end{equation}
This quantity is the ratio between the gravitational and the electromagnetic
contributions to the pulsar spin-down rate and we will see in the next section
that its present values are always less than $1$.
Note that, in terms of $\Omega$ and its derivative,
\begin{equation}
n_{em}={{\Omega~\ddot{\Omega}_{em}}\over{\dot{\Omega}\dot{\Omega}_{em}}}
\label{nem}
\end{equation}
Inverting Eq.(\ref{n1}),
\begin{equation}
Y(n)={{n-n_{em}}\over{5-n}}
\label{yn}
\end{equation}
which implies $n_{em}\le n <5$.
In Fig.(\ref{fig1}) we have plotted the function $ Y(n)$
for different values of $ n_{em}$ in the range $ [1,3].$ 
As we expect, for each $n_{em}$, $Y(n)=0$ when $n=n_{em}$ (i.e. no gravitational torque) while
$Y(n)\rightarrow +\infty $ for $n\rightarrow 5$ (i.e. negligible electromagnetic
torque).
We see that a braking index less than 3 can be obtained combining
the effects of the electromagnetic and gravitational torques. For instance,
for the Crab pulsar we have $ n=2.51$ and if we assume, 
say, $ n_{em}=2.4$, then it immediately follows $Y\simeq 0.04$, that is the
gravitational radiation contribution to the spin-down rate is about $4\% $ of the 
electromagnetic one. On the other hand, if $n_{em}=1$, then $Y\simeq 0.6$. 

The spin-down of a pulsar in which both the electromagnetic and gravitational
torques are acting, Eq.(\ref{omp}), can be expressed as
\begin{equation}
\dot{\Omega}=K_{eff}(\Omega)\times \Omega^{n(\Omega)}
\end{equation}
where both the ``effective'' $ K_{eff}$ and $ n$ are function of the angular velocity
$ \Omega .$ As a consequence of its dependence  
on $ \Omega $, the braking index $ n$ is not constant in time. 
In the next section we will derive the time evolution of $n$.
In all subsequent calculations we will assume the pulsar
momentum of inertia being equal to the canonical value $I=10^{38}~kg~m^2$.

From Eq.(\ref{ompgw}) we can express the ellipticity of the pulsar as
\begin{equation}
\epsilon=1.9\times 10^5\sqrt{P^3\dot{P}_{gw}}
\label{ep}
\end{equation}
Combining Eqs.(\ref{ompgw},\ref{ipsi}) we can re-write Eq.(\ref{ep}) as a function 
of the observed pulsar period and its derivative, and of
the ratio $Y$ between the gravitational and electromagnetic spin-down rates:
\begin{equation}
\epsilon=1.9\times 10^5\sqrt{P^3\dot{P}{Y(n)\over{1+Y(n)}}}=
7.55\times 10^6\sqrt{{|\dot{\Omega}|\over{\Omega^5}}{Y(n)\over{1+Y(n)}}}
\label{epy}
\end{equation}
Eq.(\ref{epy}) plays a basic role in the determination of the upper limits
of the pulsars ellipticity.
 
\section{Pulsar parameters evolution}
\label{evol}  
Differentiating Eq.(\ref{ipsi}) we easily find 
\begin{equation}
\dot{Y}=(5-n_{em})Y{\dot{\Omega}\over \Omega}
\label{doty}
\end{equation}
and then
\begin{equation}
Y(\Omega)=Y_0\left({\Omega \over \Omega_0}\right)^{5-n_{em}}
\label{yom}
\end{equation}
where $Y_0={{n_0-n_{em}}\over{5-n_0}}$, being $n_0$ the observed braking index
(i.e. its present value).
We use this solution to solve the equation for the time evolution of the pulsar
angular velocity. To this purpose, let us write Eq.(\ref{omp}) in the form
\begin{equation}
\dot{\Omega}=K_{gw}\Omega^5\left({1+Y(\Omega)\over{Y(\Omega)}}\right)
\label{omp2}
\end{equation}
We numerically solve this differential equation, fixing the pulsar age
$t_{age}$ and finding a solution depending on
three unknowns: the ellipticity $\epsilon$ (contained in $K_{gw}$),
$n_{em}$ (on which $Y(\Omega)$ depends) and the initial angular 
velocity $\Omega_i$ ($\Omega_i=\Omega(t_{age})$). 
These quantities must be
chosen in order to verify two conditions: those given by $\Omega(t=0)=
\Omega_0$ and by Eq.(\ref{epy}), calculated for the values of the parameters 
at $t=0$\footnote{In fact, as we assume $K_{gw}=const$, the value of $\epsilon $
coming from Eq.(\ref{epy}) is not dependent on the particular time $t$
considered.}: 
\begin{equation}
\epsilon=7.55\times 10^6\sqrt{{|\dot{\Omega}_0|\over{\Omega^5_0}}{Y_0\over{1+Y_0}}}
\label{epy0}
\end{equation}
This gives us a range of acceptable values for the unknowns.
The solution we are searching for is that corresponding to the 
smallest possible value of electromagnetic braking index $n_{em}$, that is to the 
greatest value
of the ellipticity $\epsilon$\footnote{From Eq.(\ref{epy0}), 
the greater is the difference between $n_{em}$
and $n_0$ and the greater is the ellipticity.}. Once we have determined the 
function $\Omega(t)$ for a given pulsar, we can use Eq.(\ref{n1}) and find the time
evolution of the braking index $n(t)$.

In some cases, we can further reduce the range of
variation of $n_{em}$ considering the pulsar age, in the following way.
Let us consider the ratio $R$ between $\dot{\Omega}_{em}$ and the total spin-down rate,
given by Eq.(\ref{omp2}), or equivalently, by
\begin{equation}
\dot{\Omega}=K_{em}\Omega^{n_{em}}\left(1+Y(\Omega)\right)
\label{omptot}
\end{equation}
We have
\begin{equation}
R(\Omega)={\dot{\Omega}_{em}\over{\dot{\Omega}}}={1\over{1+Y(\Omega)}}
\label{rap}
\end{equation}
This quantity varies monotonically between the two limits
$R_{inf}=R(\Omega_i)$ and $R_{sup}=R(\Omega_0)={{5-n_0}\over{5-n_{em}}}\le 1$. 
The condition $\dot{\Omega}_{em}\le R_{sup}\dot{\Omega}$ implies
\begin{equation}
R_{sup}\ge {t_{age}\over \tau_{em}}
\label{rsup}
\end{equation}
with $\tau_{em}$ given by Eq.(\ref{tem}).
We choose the range of $n_{em}$ in order to satisfy Eq.(\ref{rsup}).
In this way, with the exception of the Vela pulsar, we can increase the lower limit
of $n_{em}$ then reducing the upper value of $\epsilon$. 

In the following we consider individually the four
pulsars with measured braking index, applying to each of them the procedure
described.

\subsection{Crab}
\label{crab}
The Crab pulsar present parameters are $n_{0}=2.51\pm 0.01$, 
$\Omega_{0}=188.12~s^{-1}$ and $\dot{\Omega}_{0}=-2.42\times 10^{-9}s^{-2}$, 
while its age is $t_{age}\simeq 945~yr$.
The initial rotation period of the Crab is not exactly known but a value 
$P_i\in[10,20]~ms$ is 
usually assumed. 
By solving Eq.(\ref{omp2}), under the various conditions previously described, 
we find $n_{em}\ge 1.7$,
$\epsilon\le 3.8\times 10^{-4}$, $\Omega_i\le 394~s^{-1}$ ($P_i\ge 16~ms$). 
A more tightening limit can be obtained considering that  
the observed braking index appears to be
constant within $0.5\%$ in more than twenty years of observations, thus implying
$|\dot{n}|<5\times 10^{-4}~yr^{-1}$ (Lyne et al. \cite{lyne1}). This relation is 
satisfied if $n_{em}\ge 2.05$ and correspondingly we find 
$\epsilon \le 3\cdot 10^{-4}$, 
$2.5$ times lower than the maximum ellipticity calculated under the 
hypothesis that the pulsar spin-down is due only to the emission of gravitational 
waves (Haensel \cite{haens}), (Giazzotto et al. \cite{giazo}),  
$\Omega_i\le 364~s^{-1}$ ($P_i\ge 17.2~ms$) and $Y_0\le 0.18$. 
 
\subsection{Vela}
\label{vela}
The Vela pulsar is characterized, at present, by $n_0=1.4\pm 0.2$, $\Omega_{0}=70.36~s^{-1}$
and $\dot{\Omega}_{0}=-9.85\times 10^{-11}~s^{-2}$. Its age is not well
determined but estimates, based on the position offset and the proper motion
of the pulsar, give values in the range $[18,31]\times 10^3~yr$ 
(Aschenbach et al. \cite{asche}).  
Assuming $t_{age}=1.8\times 10^4~yr$, we have $n_{em}\ge 1.27$,
$\epsilon \le 3.3\times 10^{-4}$,
about $5.5$ times lower than the limit relative to the case
of gravitational torque alone, 
$\Omega_i\le 391~s^{-1}$ ($P_i\ge 16.1~ms$) and $Y_0\le 0.036$. If 
$t_{age}=3.1\times 10^4~yr$ we find much more tightening limits:
$n_{em}\ge 1.399$, $\epsilon \le 3\times 10^{-5}$,
about $60$ times lower than the limit relative to the case
of gravitational braking alone,
together with 
$\Omega_i\le 594~s^{-1}$ ($P_i\ge 10.6~ms$) and $Y_0\le 2.78\cdot 10^{-4}$. 
\subsection{PSR 0540-69}
\label{psr05}
The pulsar $PSR~ 0540-69$ is characterized by $n_0=2.28\pm 0.02$, $\Omega_{0}=124.66
~s^{-1}$ and $\dot{\Omega}_{0}=-1.185\times 10^{-9}~s^{-2}$. We do not know
exactly
its age which, however, should be of about $900~yr$ 
(Muslimov $\&$ Page \cite{musli}). 
We find $n_{em}\ge 2.21$, $\epsilon \le 2.4\times 10^{-4}$, which is $6.3$ times
lower than the ``gravitational'' limit,  
$\Omega_i\le 175~s^{-1}$ ($P_i\ge 36~ms$) and $Y_0\le 0.026$. 
\subsection{PSR 1509-58}
\label{psr1509}
The pulsar $PSR~1509-58$ parameters are: $n_0=2.837\pm 0.001$, $\Omega_{0}=41.7~s^{-1}$ and
$ \dot{\Omega}_{0}=-4.25\times 10^{-10}~s^{-2}$. If the pulsar birth is identified
with the supernova event $AD~185$, then its age is $t_{age}\simeq 1800~yr$ 
(Thorsett \cite{thors}), (Muslimov $\&$ Page \cite{musli}). On the other hand its ``electromagnetic'' age is
$\tau_{em}\le 1700~yr$. The origin of this discrepancy has not been 
understood yet. So, our model cannot be applied because the pulsar age
resulting from the action of both the gravitational and electromagnetic torques
is always less than $\tau_{em}$ and cannot match $t_{age}$.
\begin{figure}
\resizebox{\hsize}{!}{\includegraphics{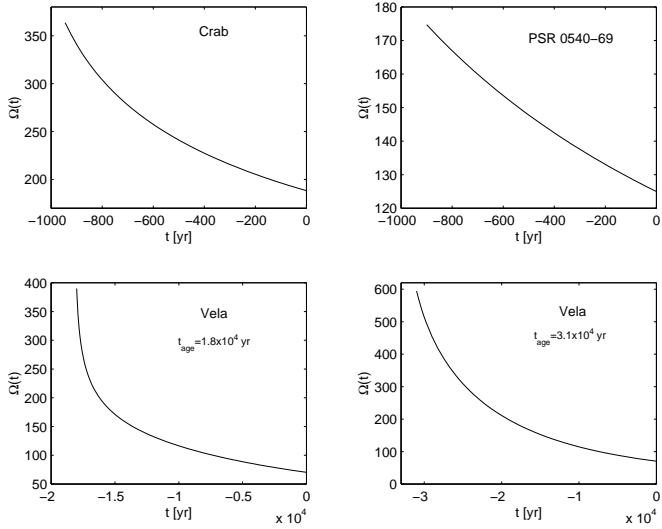}}
\caption{The angular velocity $\Omega(t)$ is plotted as a function of time,
assuming $n_{em}$ equal to its lower limit. 
For the Vela pulsar age both the lower and the upper limit have been considered.}
\label{variom}
\end{figure} 
\begin{figure}
\resizebox{\hsize}{!}{\includegraphics{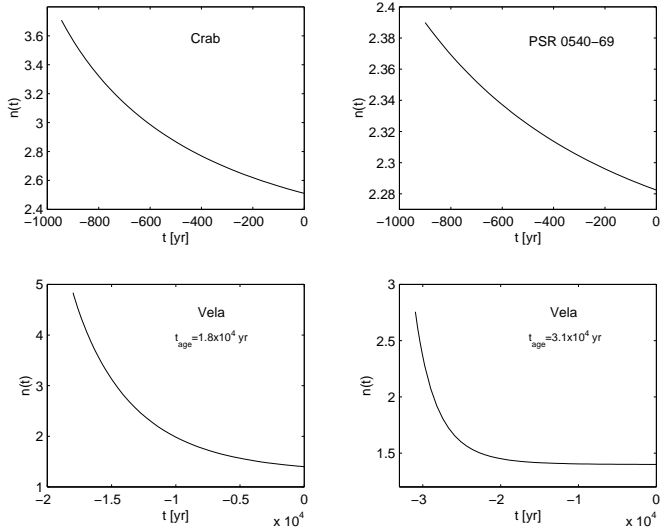}}
\caption{The braking index $n$ is plotted as a function of time, 
assuming $n_{em}$ equal to its lower limit. For the Vela
pulsar age both the lower and the upper limit have been considered.}
\label{nt}
\end{figure} 

We have plotted the time evolution of the pulsars angular
veocity $\Omega(t)$ in Fig.(\ref{variom}) and that of the braking index
$n(t)$ in Fig.(\ref{nt}). In all cases $n(t)\rightarrow n_{em}$ asymptotically,
because the gravitational contribution to the spin-down rate decreases faster than
the electromagnetic one, see Eq.(\ref{ipsi}).We note that for the Vela pulsar, 
if $t_{age}=1.8\cdot 10^4~yr$, the initial value of the braking index $n$ could be
as high as $4.8$. This implies that $Y_i\equiv Y(\Omega_i)\le 22$, so that at the 
beginning the 
spin-down could have been dominated by the gravitational braking. From the figure we
see that $Y$ would have become less than $1$ about $3000~yr$ after the birth. Also
for the Crab the gravitational braking could have been dominant at the birth
($Y_i\le 1.28$).

Our results, together with some pulsars parameters, are
summarized in Tab.(\ref{tabl}).

\section{Emission of gravitational waves}
\label{graw}
From the calculated upper limits on the ellipticity of pulsars we can
very easily find new upper limits also on the amplitude of the gravitational
signals emitted by them. For a deformed neutron star rotating around a
principal axis of inertia, the waveforms corresponding to the two independent
polarization states are, in the quadrupole formalism, 
\begin{equation}
h_+={h_0\over 2}\cos{2\Omega t}\left(1+\cos^2{\theta}\right)
\label{accapi}
\end{equation}
\begin{equation}
h_{\times}=h_0\sin{2\Omega t}\cos{\theta}
\label{accape}
\end{equation}
where $\theta$ is the angle between the rotation axis of the star and the line
of sight and the amplitude $h_0$ is given by
$$
h_0=1.05\times 10^{-27}\left({I\over{10^{38}kg~m^2}}\right)\left({10~kpc\over r}
\right)\times 
$$
\begin{equation}
~~~~\times \left({\epsilon\over{10^{-6}}}\right)\left({\nu\over{100~Hz}}\right)^2
\label{h0}
\end{equation}
being $\nu={\Omega\over{\pi}}$ the frequency of the wave, equal to two times
the rotation frequency of the star. For completeness, we remember that the
power emitted through gravitational waves is 
\begin{equation}
{dE\over{dt}}=-{32\over{5}}{G\over{c^5}}I^2\epsilon^2\Omega^6
\label{dedt}
\end{equation}
In Tab.(\ref{tabl}) we have reported, among the other quantities, the 
maximum amplitude of the gravitational waves emitted from the pulsars with
known braking index, calculated by the use of Eq.(\ref{h0}), assuming 
$I=10^{38}~kg~m^2$.
\begin{table*}$$
\begin{tabular}{|c|c|c|c|c|c|c|c|} \hline
$~~$&$ r~[kpc] $&$ P_0~[ms] $&$ \nu~[Hz] $&$ P_{i,min}~[ms] $&$ Y_{0,max} $&$ 
\epsilon_{max} $&$ h_{max} $\\ \hline
$ Crab $&$ 2 $&$ 33.4 $&$ 59.9 $&$ 17.2 $&$ 0.18 $&$ 3\times 10^{-4} 
$&$ 5.5\cdot 10^{-25}$ \\ \hline
$ Vela $&$ 0.5 $&$ 89.3 $&$ 22.4 $&$ 16.1^a $&$ 0.036 $&$ 3.3\times 10^{-4} 
$&$ 3.5\times 10^{-25}$ \\ \hline
$      $&$   $&$      $&$      $&$ 10.6^b $&$ 0.00028 $&$ 3\times 10^{-5} 
$&$ 3.2\times 10^{-26}$ \\ \hline
$PSR~0540-69 $&$ 49.4 $&$ 50.4 $&$ 39.7 $&$ 36 $&$ 0.026 $&$ 2.4\times 10^{-4} 
$&$ 8.1\times 10^{-27}$ \\ \hline
\end{tabular}$$
\caption{Parameters of pulsars: $r$ is the pulsar
distance, $P_0$ is the measured
period, $\nu $ is the frequency of the gravitational waves emitted (twice the
orbital frequency), $P_{i,min}$ is the minimum initial period, 
$Y_{0,max}$ is the maximum present value of the ratio between the gravitational
and the electromagnetic contributions to the pulsar spin-down rate, 
$\epsilon_{max}$ and $h_{max}$ are, respectively, the maximum ellipticity and
the maximum amplitude of the gravitational waves emitted. For the Vela pulsar
two ages, $t_{age}=1.8\times 10^4~yr$ (a) and $t_{age}=3.1\times 10^4~yr$ (b)
have been considered.}
\label{tabl}
\end{table*}

\section{Conclusions}
\label{concl}
Usually, the upper limits on pulsars ellipticity are calculated assuming that
pulsar spin-down is due only to the emission of gravitational waves. It is
well known that this assumption contradicts the measure of the braking index,
which has been obtained for a small sample of pulsars, with values always
less than 3.

In this paper we show that considering the simultaneous action, in a pulsar,
of a gravitational torque and an electromagnetic one, the observed braking
indexes can be recovered. The gravitational torque is always a small 
correction to the total torque acting on the pulsar. A consequence of the 
model is that the braking index is not constant in time. We calculate the
time evolution of the pulsars angular velocities and braking indexes and
find more tightening upper limits to their ellipticity.
These new limits are between $2.5$ and $60$ times lower than
the old one. New upper limits on the amplitude of the gravitational waves
emitted immediately follow. 
\begin{acknowledgements}
I would like to thank the anonymous referee for his remarks and his helpful 
suggestions.
\end{acknowledgements}

\end{document}